\documentclass[journal=nalefd,manuscript=letter,layout=twocolumn]{achemso}
\setkeys{acs}{maxauthors = 0}
\setkeys{acs}{articletitle = false}

\usepackage[version=3]{mhchem} 
\usepackage{graphicx}
\usepackage{dcolumn}
\usepackage{bm}
\usepackage[mathlines]{lineno}
\usepackage{amssymb}
\usepackage{amsmath}
\usepackage{natbib}
\usepackage{soul,color}
\usepackage{xcolor,colortbl}
\usepackage{multirow}
\usepackage{color}%
\newcommand{\modR}[1]{{#1}}

\newcommand*{\onlinecite}[1]{[\hspace{-1 ex} \nocite{#1}\citenum{#1}]}

\makeatletter
\renewcommand*{\@fnsymbol}[1]{\ensuremath{\ifcase#1\or \dagger\or *\or \ddagger\or
\mathsection\or \mathparagraph\or \|\or **\or \dagger\dagger \or
\ddagger\ddagger \else\@ctrerr\fi}} \makeatother

\title{Effect of Net Charge on the Relative Stability of
       2D Boron Allotropes}

\author{Dan~Liu}
\affiliation{Physics and Astronomy Department,
             Michigan State University,
             East Lansing, Michigan 48824, USA}

\author{David Tom\'{a}nek}
\email
            {tomanek@pa.msu.edu}%
\affiliation{Physics and Astronomy Department,
             Michigan State University,
             East Lansing, Michigan 48824, USA}

\date{\today} 

\abbreviations{DFT, 2D}


\begin{document}


\begin{abstract}
We study the effect of electron doping on the bonding character
and stability of two-dimensional (2D) structures of elemental
boron, called borophene, which is known to form many stable
allotropes. Our {\em ab initio} calculations for the neutral
system reveal previously unknown stable 2D $\epsilon$-B and
$\omega$-B structures. We find that the chemical bonding
characteristic in this and other boron structures is strongly
affected by extra charge. Beyond a critical degree of electron
doping, the most stable allotrope changes from $\epsilon$-B to a
buckled honeycomb structure. Additional electron doping, mimicking
a transformation of boron to carbon, causes a gradual decrease in
the degree of buckling of the honeycomb lattice that can be
interpreted as piezoelectric response. Net electron doping can be
achieved by placing borophene in direct contact with layered
electrides such as Ca$_{2}$N. We find that electron doping can be
doubled by changing from the B/Ca$_{2}$N bilayer to the
Ca$_{2}$N/B/Ca$_{2}$N sandwich geometry.
\end{abstract}

%


Identifying the most stable allotrope of a given compound is one
of the key problems in Physics and Chemistry. Whereas charge
neutral systems have attracted most attention, notable exceptions
are low-dimensional systems that can be charged by doping in
specific bulk geometries. Electron doping induced by Li
intercalation does not affect the honeycomb structure of
two-dimensional (2D) graphene layers in graphite intercalation
compounds (GICs)~\cite{Dresselhaus81}, but does change the
equilibrium structure of MoS$_2$ monolayers from the 2H to the 1T
allotrope~\cite{Haering83}. Even though the effect of excess
charge on chemical bonding and equilibrium geometry should be
general, we expect the most drastic changes to occur in structures
of elemental boron that is notorious for its many stable
allotropes~\cite{{Cheng12},{Ihsan05}}.
What appears as frustrated bonding in %
mostly metallic 2D structures of elemental boron, called
borophene, reflects the inability of the element to follow the
octet rule of bonding due to its electron deficiency. We speculate
that both the electron deficiency and thus the chemical bonding
characteristic may be modified by placing a nonzero charge on B
atoms. In that case, the most stable borophene structure may
differ from a puckered triangular
lattice with monatomic vacancies~\cite{%
{Pantelides05},{Alexander06},{Mannix15},{Kehui16},{Yakobsonrev17}}
or equally stable irregular
structures~\cite{{Zhou13},{Zhou14},{Ma16}} identified in the
neutral system. Indeed, placing a net charge of ${\lesssim}1$~e
per atom on borophene, provided by an Al(111)
substrate~\cite{Wu18}, or intercalating Mg ions in-between
borophene layers in the MgB$_2$ compound~\cite{MgB2bonding01},
changes the most stable allotrope to a very different honeycomb
lattice. Structural changes induced by charging may significantly
modify the electronic structure, turning semiconducting 2H-MoS$_2$
to metallic 1T'-MoS$_2$ locally~\cite{Chhowalla14} and undoped
borophene to honeycomb lattices in
diborides~\cite{{Matkovich77},{Stefano06}} including MgB$_2$,
which displays superconducting behavior~\cite{Akimitsu01}.

In this study, we explore the effect of net charge on the bonding
character and structural stability of 2D allotropes of boron. Our
{\em ab initio} calculations for the neutral system reveal a
previously unknown stable 2D $\epsilon$-B structure with a 0.2~eV
wide fundamental band gap. We find that the chemical bonding
characteristic in this and other boron structures is strongly
affected by extra charge, including a $23$\% lattice constant
change in $\epsilon$-B, induced by changing the net charge from
0.25 holes to 0.25 electrons per B~atom. Beyond a critical degree
of doping near 0.5~electrons/atom, the most stable allotrope
changes from $\epsilon$-B to a buckled honeycomb structure.
Additional electron doping, mimicking a transformation of boron to
carbon, causes a gradual decrease in the degree of buckling of the
honeycomb lattice that can be interpreted as piezoelectric
response. We propose that net electron doping can easily be
achieved by placing borophene in direct contact with layered
electrides such as Ca$_{2}$N. In this system, we find that
electron doping of borophene can be doubled by changing from the
B/Ca$_{2}$N bilayer to the Ca$_{2}$N/B/Ca$_{2}$N sandwich
geometry.

As mentioned above, the vast number of stable neutral 2D borophene
allotropes, including $\alpha-$B, $\beta-$B, $\gamma-$B,
$\delta-$B and $\eta-$B
structures~\cite{{Pantelides05},{Alexander06}}, reflects a
frustrated bonding character of electron-deficient boron. This
element may engage its three valence electrons in only three
covalent bonds, resulting in an electron sextet instead of the
desirable octet noble-gas configuration. To partly compensate for
lack of electrons in the octet configuration, boron atoms often
prefer increasing their coordination to six nearest neighbors in
the triangular lattice. This is essentially equivalent to adding
electrons~\cite{{Sohrab07},{Sohrab09}} to the individual atoms
that are held together by pure three-center bonding. On the other
hand, the stability of the triangular lattice is often enhanced by
removing atoms and forming hexagon-shaped monatomic vacancies.
This process somehow mimics subtracting electrons while the
structure locally converts to a honeycomb lattice, where atoms are
held together by two-center bonding. The competition between
two-center and three-center bonding has been used to identify the
optimum concentration of hexagonal vacancies in the neutral
triangular lattice~\cite{{Sohrab10},{Sohrab07},{Eluvathingal17}},
but likely controls also less common neutral structures with four-
and five-fold coordinated boron atoms~\cite{{Zhou13},{Zhou14}}.
\modR{%
Biasing this competition by net charge has been shown to affect
the fraction of hexagonal vacancies in triangular borophene
lattices~\cite{Tarkowski18}. Even though doping of B layers has
been linked to unexpected superconducting behavior of MgB$_2$ and
found useful to modulate CO$_{2}$ capture~\cite{Smith17}, no
systematic attention has been paid to the possibility of
deliberately changing the bonding and the
equilibrium structure of boron by excess charge. %
}

A viable possibility to significantly dope 2D structures of
elemental boron by electrons is to place them in direct contact
with electrides including Ca$_{2}$N~\cite{{Scott16},{Hideo13}} and
Y$_{2}$C~\cite{Hideo14}. In these highly electronegative systems
with a layered structure, regions of large electron density are
found in-between the layers. The charge density in this electron
layer amounts to one electron per formula unit in Ca$_{2}$N and
two electrons per formula unit in Y$_{2}$C. The possibility of
exfoliation down to a monolayer~\cite{Scott16} has been
demonstrated in Ca$_{2}$N, so that assembly of various vertical
heterostructures is possible.

\section*{Results}

\subsection*{Allotropes of neutral and doped borophene}


Inspired by the honeycomb structure of the negatively doped boron
sublattice found in diborides including MgB$_2$, we started our
investigation of electron doped borophene structures with the
honeycomb lattice. To provide substantial configuration freedom
for the lattice structure, we consider a superlattice with 32
atoms per unit cell and subject all atoms in the supercell to
random distortion. To study the effect of doping on the
equilibrium structure and the chemical nature of bonding, we
changed the degree of doping gradually and optimized each system
using the conjugate gradient optimization method.

\begin{figure}[t]
\includegraphics[width=1.0\columnwidth]{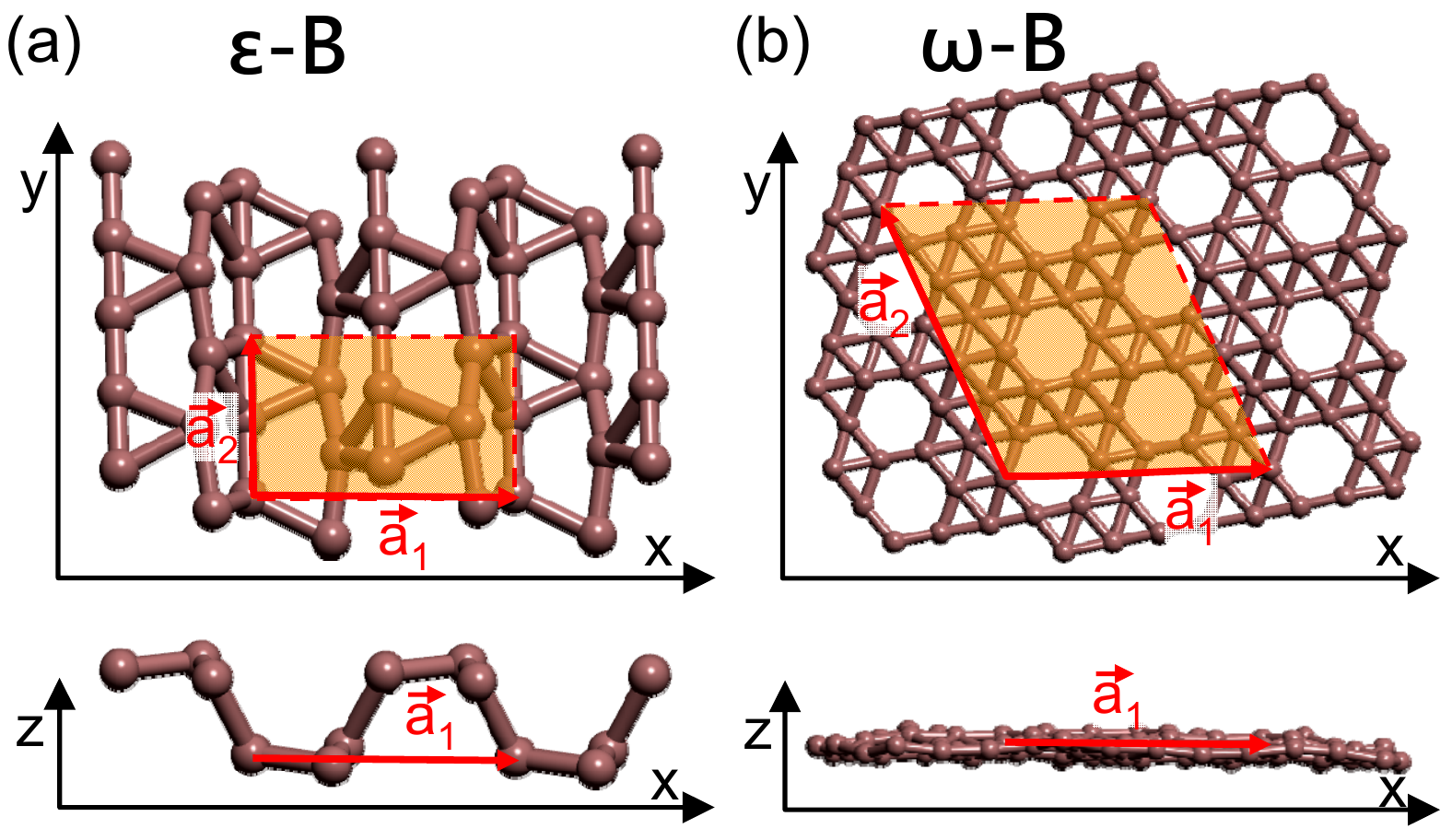}
\caption{%
Previously unexplored neutral borophene allotropes formed by
spontaneous conversion of an artificial honeycomb lattice: %
(a) $\epsilon$-B and 
(b) $\omega$-B.      
The structures are shown in top and side view. The lattice vectors
$\vec{a}_1$ and $\vec{a}_2$, shown in red, delimit the highlighted
primitive unit cells. %
\label{fig1}}
\end{figure}

\begin{table}[b!]
\caption{%
Cohesive energy $E_{coh}$ of selected neutral borophene
allotropes,
obtained using DFT-PBE calculations. $n$ is the number of boron
atoms per unit cell and $Z$ is the coordination number of
individual atoms in the unit cell.%
}%
\begin{tabular}{lccc} %
\hline \hline
   \textrm{Allotrope} %
 & \textrm{$E_{coh}$~(eV/atom)} %
 & \textrm{$n$} %
 & \textrm{$Z$} %
 \\
\hline%
  {$\epsilon$-B} %
& {$5.699^a$} %
& {8} %
& {4}%
\\
  {$\omega$-B} %
& {$5.680^a$} %
& {32} %
& {4, 5, 6}%
\\
%
  {$\alpha'$-B} %
& {$5.706^a$} %
& {8} %
& {5, 6} %
\\
   %
& {$5.762^b$} %
& {8} %
& {5, 6} %
\\
  {$\beta_{12}$-B} %
& {$5.712^b$} %
& {5} %
& {3, 5, 6}%
\\
& {$6.23^c$} %
& {5} %
& {3, 5, 6}%
\\
  {$\delta_{6}$-B} %
& {$5.662^b$} %
& {4} %
& {6}%
\\
  {$\chi_{3}$-B} %
& {$5.723^b$} %
& {4} %
& {3, 5}%
\\
& {$6.19^c$} %
& {4} %
& {3, 5}%
\\
\hline \hline %
$^a$ Present work. \\
$^b$ Reference \onlinecite{Cheng12}.\\
$^c$ Reference \onlinecite{Kehui16}.\\
\end{tabular}
\label{table1}
\end{table}

\begin{figure}[t]
\includegraphics[width=1.0\columnwidth]{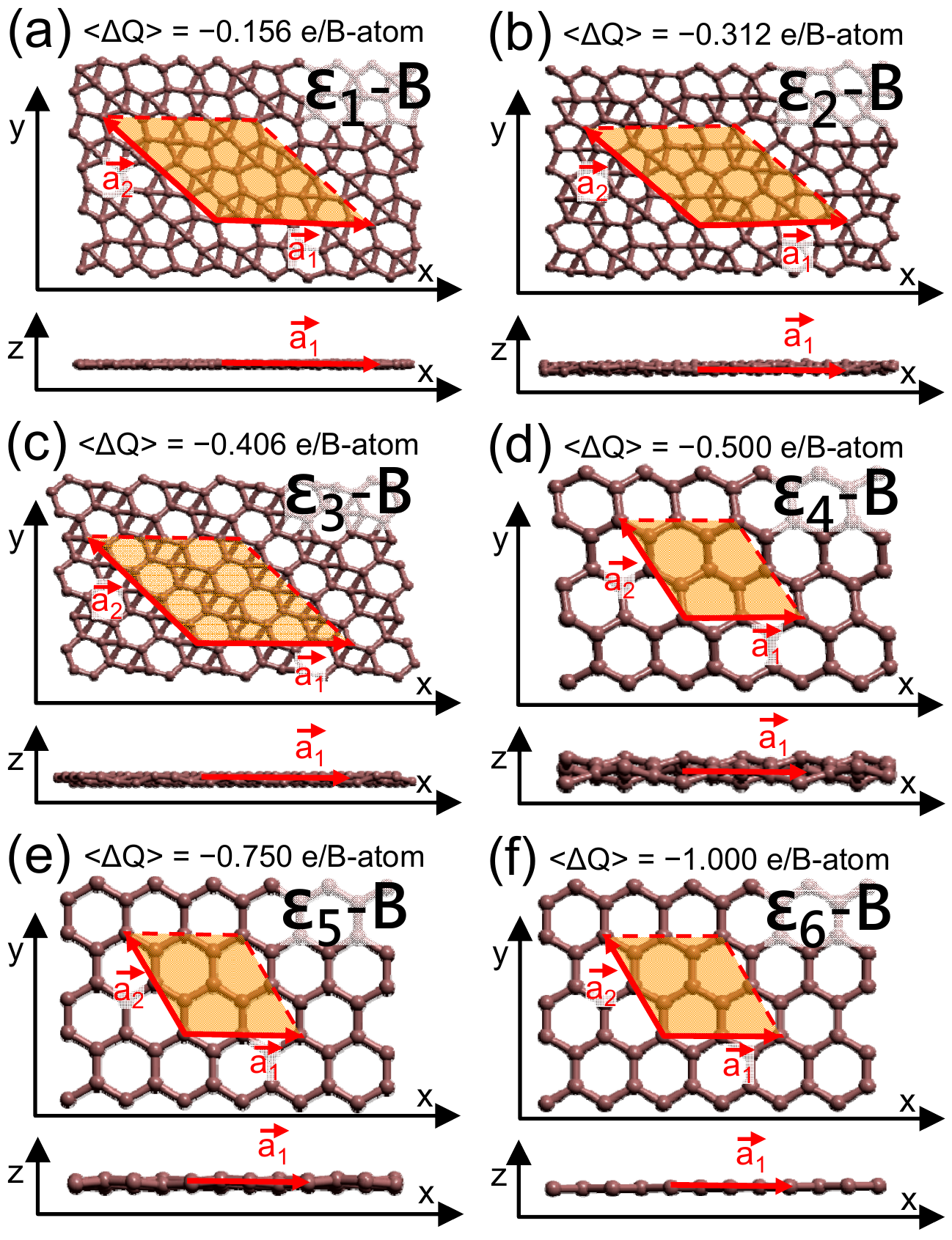}
\caption{Electron doped 2D borophene structures obtained by
optimizing a distorted boron honeycomb superlattice with 32 atoms
per unit cell. The average excess charge $\langle{\Delta}Q\rangle$
per boron atom, specified in the panels, increases from (a)
$\epsilon_1$-B to (f) $\epsilon_6$-B. The structures are shown in
top and side view. The lattice vectors $\vec{a}_1$ and
$\vec{a}_2$, shown in
red, delimit the highlighted unit cells. %
\label{fig2}}
\end{figure}

Starting with no excess charge, we found the neutral honeycomb
structure to be unstable and to convert to rather stable
allotropes that have not been reported previously. The first
allotrope, shown in Fig.~\ref{fig1}(a) and called $\epsilon$-B, is
only 7~meV/atom less stable than the most stable $\alpha'$-B
phase~\cite{Cheng12} and displays a very uncommon morphology with
triangles and pentagons, %
\modR{quite distinct from the well-documented class of triangular
lattices with vacancies. %
}%
The structure shown in Fig.~\ref{fig1}(b), which we call
$\omega$-B, has a very similar morphology to known allotropes
containing
triangles and hexagons only~\cite{%
{Pantelides05},{Alexander06},{Mannix15},{Kehui16}}, but is
$26$~meV/atom less stable than $\alpha'$-B. $\epsilon$-B and
$\omega$-B add to the large number of known allotropes, and we
expect many more to follow.

Both $\epsilon$-B and $\omega$-B are buckled. Still, we could
locate a locally stable, flat counterpart of $\omega$-B, which is
7~meV/atom less stable than $\omega$-B. The cohesive energies of
neutral borophene allotropes are compared in Table~\ref{table1}.

Next, we added extra 5, 10, 13, 16, 24 and 32 electrons to the
32-atom unit cell with an initial honeycomb arrangements and
optimized the geometry. %
\modR{%
We found the optimum geometries not to depend on initial
deformations imposed on the starting structure, which may not be
reachable in molecular dynamics trajectories.
}%
The optimum boron structures labelled $\epsilon_1-\epsilon_6$ and
their average doping levels $\langle{\Delta}Q\rangle$ are
displayed in Figs.~\ref{fig2}(a)-\ref{fig2}(f). Optimum structures
for different doping levels are discussed in the Supporting
Information (SI). We find all these structures to differ
significantly from the neutral $\epsilon$-B and $\omega$-B
structures in Fig.~\ref{fig1}, which have been optimized in the
same way.

\begin{figure*}[t]
\includegraphics[width=1.6\columnwidth]{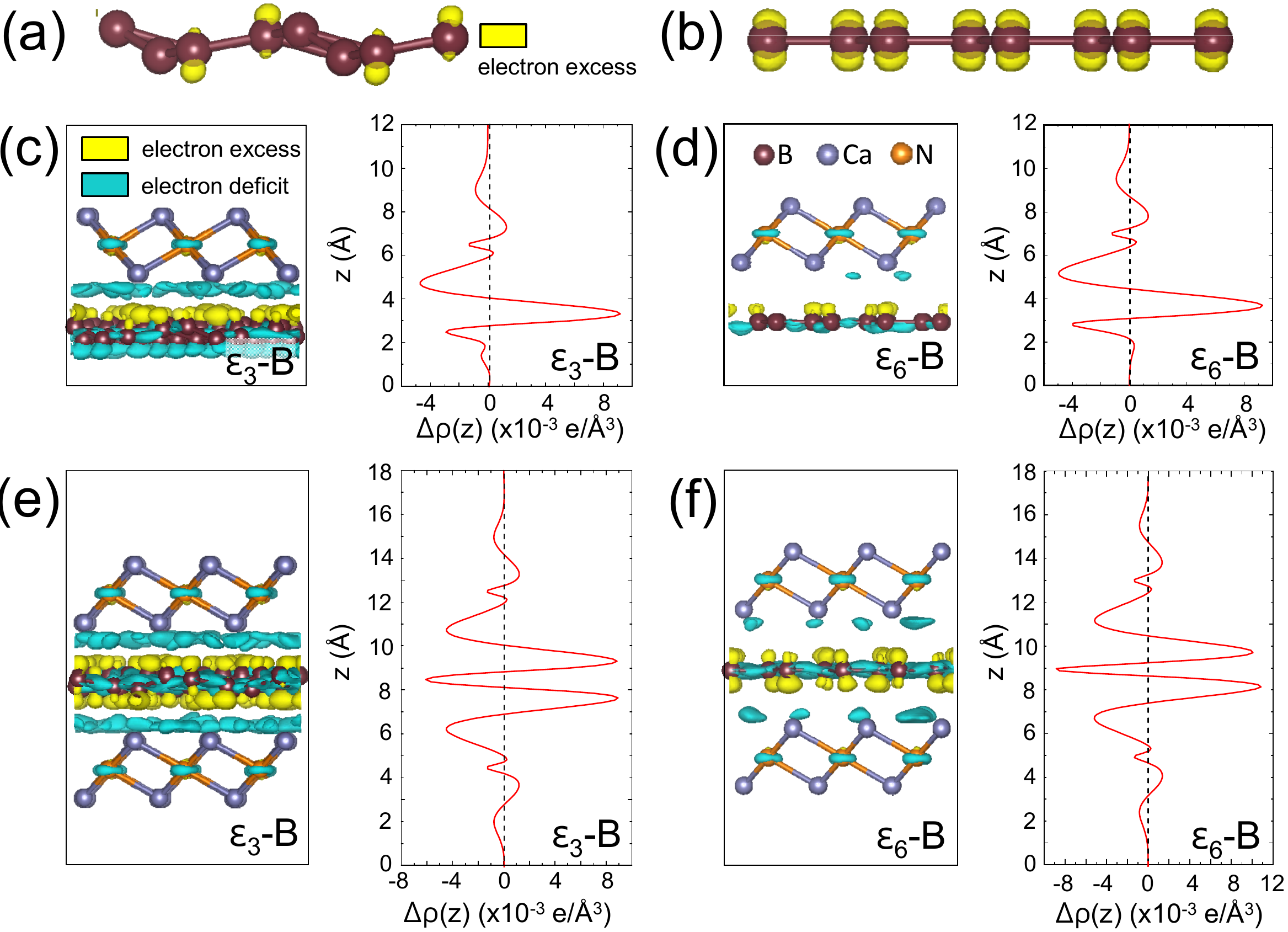}
\caption{ %
Charge redistribution in 2D borophene layers induced by electron
doping or by contact with monolayers of Ca$_{2}$N electride.
Charge density difference ${\Delta}\rho$ caused by placing an
excess charge %
(a) $\langle{\Delta}Q\rangle=-0.406$~e/atom on $\epsilon_3$-B and %
(b) $\langle{\Delta}Q\rangle=-1.0$~e/atom on $\epsilon_6$-B. %
Charge density redistribution
${\Delta}\rho=\rho$(B/Ca$_{2}$N)$-\rho$(B)$-\sum\rho$(Ca$_{2}$N)
in the bilayer structures %
(c) $\epsilon_3$-B/Ca$_{2}$N and %
(d) $\epsilon_6$-B/Ca$_{2}$N %
as well as in the sandwich structures %
(e) Ca$_{2}$N/$\epsilon_3$-B/Ca$_{2}$N and %
(f) Ca$_{2}$N/$\epsilon_6$-B/Ca$_{2}$N. %
${\Delta}\rho$ is shown by isosurfaces bounding regions of
electron excess at $+7{\times}10^{-3}~\text{e}$/{\AA}$^3$ (yellow)
and electron deficiency at
$-2\times10^{-3}~\text{e}$/{\AA}$^3$ (blue). %
$\langle{\Delta}\rho(z)\rangle$ is averaged across the $x-y$ plane
of the layers.
\label{fig3}}
\end{figure*}

The structure $\epsilon_1$-B in Fig.~\ref{fig2}(a), which contains
triangles, pentagons and hexagons, is reminiscent of $\epsilon$-B,
but is completely flat. With increasing electron doping, the
density of pentagons gradually diminishes and eventually vanishes
in $\epsilon_3$-B in Fig.~\ref{fig2}(c), representing the buckled
$\chi_{3}$ phase~\cite{Cheng12}. At the same time, the density of
hexagons increases from $\epsilon_1$-B to $\epsilon_3$-B until all
other polygons are eliminated in $\epsilon_4-\epsilon_6$, shown in
Figs.~\ref{fig2}(d)-\ref{fig2}(f), as the doping level exceeds
$|\langle{\Delta}Q\rangle|{\gtrsim}~0.5$~e/atom. The buckled
honeycomb lattice of $\epsilon_4$-B gradually flattens to the
graphene-like structure of $\epsilon_6$-B with increasing electron
doping.


The interpretation of these structural changes is rather %
\modR{%
straight-forward. With one extra electron per atom, boron behaves
as $sp^2$-bonded carbon with four valence electrons, with atoms
forming the 2D graphene honeycomb lattice. This structure is
rather robust with respect to electron and hole doping, as
evidenced in GIC structures. Similarly, also doped boron should
keep its optimum honeycomb lattice structure even if the net
charge may be smaller or larger than one extra electron
per atom. %
 }%
As seen in Figs.~\ref{fig2}(d)-\ref{fig2}(f), the honeycomb
structure of borophene, with different degree of buckling, is
preferred for the excess charge $\langle{\Delta}Q\rangle$ ranging
between $-0.5$ and $-1.0$~e/atom.

To understand the change in the electronic structure that caused a
profound structural change from neutral $\epsilon$-B in
Fig.~\ref{fig1}(a) to negatively charged $\epsilon_6$-B$^-$ in
Fig.~\ref{fig2}(f), we first probed the charge density changes
${\Delta}\rho$ associated with charging. Our results for
${\Delta}\rho$ caused by placing artificially extra electrons on
free-standing borophene structures $\epsilon_3$-B and
$\epsilon_6$-B are shown in Figs.~\ref{fig3}(a)-\ref{fig3}(b).
${\Delta}\rho$ can be viewed as the crystal counterpart of the
Fukui function, and our results indicate that the extra electrons
are accommodated in $p_z$ (or $\pi$) states normal to the layer,
very similar to graphene layers.

\begin{figure}[h]
\includegraphics[width=1.0\columnwidth]{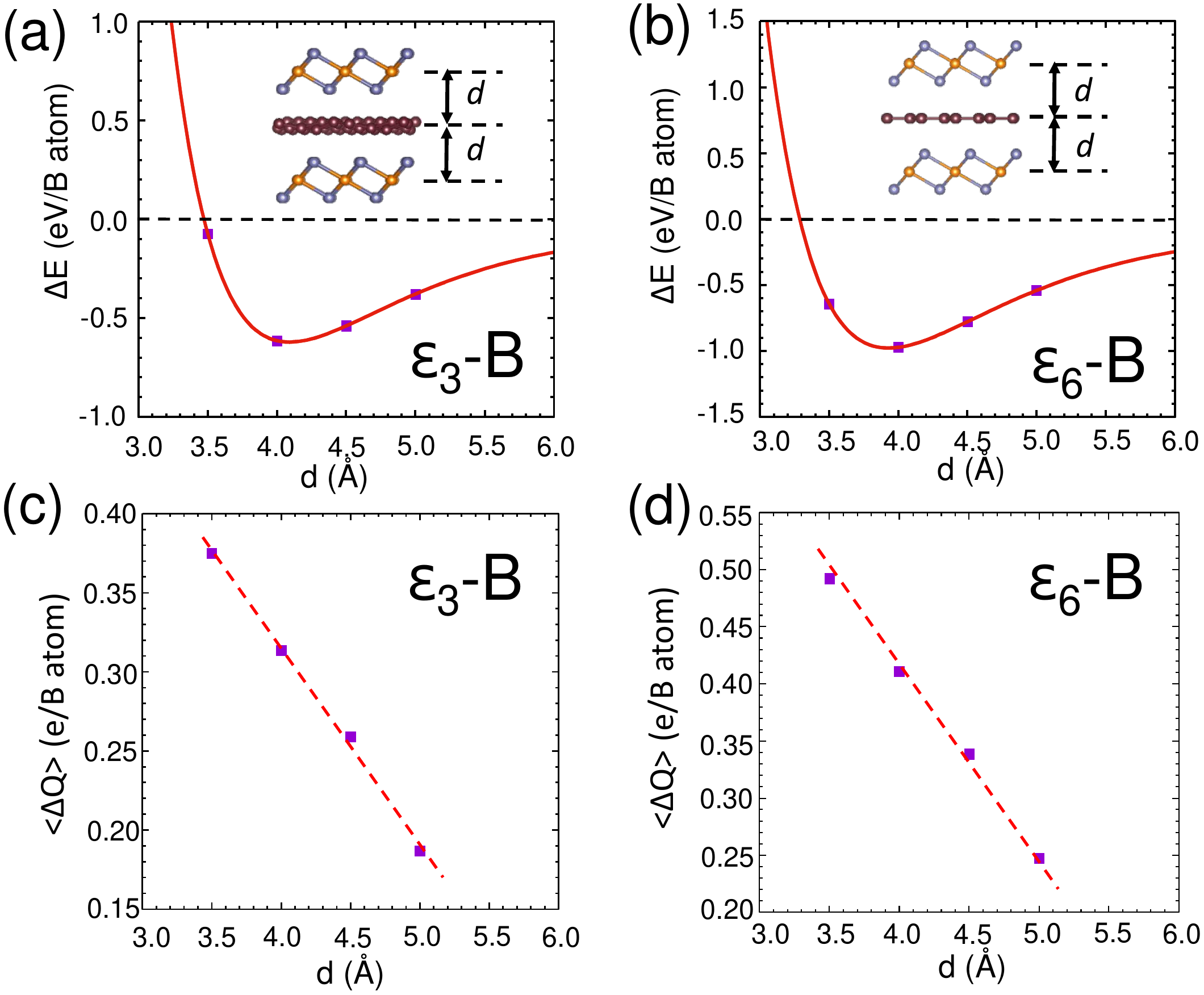}
\caption{Changes in the interlayer interaction energy ${\Delta}E$
and net average charge $\langle{\Delta}Q\rangle$ on borophene in
the Ca$_{2}$N/B/Ca$_{2}$N sandwich geometry as a function of the
interlayer distance $d$. Results for $\epsilon_3$-B in (a) and (c)
are compared to those for $\epsilon_6$-B in (b) and (d).
\label{fig4}}
\end{figure}

A realistic way to provide a high degree of electron doping, we
suggest, is to place borophene layers in contact with the
Ca$_{2}$N electride. The optimum lattice constant of the
triangular lattice of Ca$_{2}$N is $a=3.97$~{\AA}. This layered
system has the nominal configuration
[Ca$_{2}$N]$^{+}{\cdot}$e$^{-}$, with layers of Ca$_{2}$N
separated by layers of excess electrons, and can be exfoliated
down to a monolayer chemically~\cite{Scott16}. It is to be
expected that borophene will be electron doped when placed in the
region of excess electrons outside a Ca$_{2}$N monolayer. Since we
focus on general trends rather than minute details, we studied the
charge redistribution using only two prototype structures of doped
borophene, namely $\epsilon_3$-B at low- and $\epsilon_6$-B at
high-level doping, in contact with Ca$_{2}$N. To determine the
degree of electron doping in borophene caused by a contact to
Ca$_{2}$N, we inspected the charge redistribution when placing
$\epsilon_3$-B and $\epsilon_6$-B on top of a Ca$_{2}$N monolayer
or, as in a sandwich geometry, in-between Ca$_{2}$N monolayers.
The charge density differences caused by electron redistribution
in the system are shown in Figs.~\ref{fig3}(c)-\ref{fig3}(f).
Additional results for $\epsilon_3$-B and $\epsilon_6$-B on top of
a [Ca$_{2}$N]$_2$ bilayer are presented in the SI.

We should point out that incommensurate vertical heterostructures
formed of doped borophene and Ca$_{2}$N layers can not be
represented accurately in a periodic structure used in our
computational approach. In our calculation, we used the optimum
interlayer distance $d=4.0$~{\AA} in agreement with our results in
Figs.~\ref{fig4}(a) and \ref{fig4}(b). We furthermore matched %
$1{\times}3$ $\epsilon_3$-B supercells with %
$3{\times}7$ Ca$_{2}$N supercells %
in the $\epsilon_3$-B/Ca$_{2}$N superstructure and
primitive unit cells %
of $\epsilon_6$-B  with %
$3{\times}3$ supercells of Ca$_{2}$N  %
in the $\epsilon_6$-B/Ca$_{2}$N superstructure.
The remaining lattice mismatch of ${\lesssim}2$\%
was accommodated by averaging the lattice constants of borophene
and Ca$_{2}$N. Due to this minor lattice distortion, results
presented in Figs.~\ref{fig3}(c)-\ref{fig3}(f) may differ to a
small degree from the charge redistribution in an incommensurate
structure. Comparing results for the bilayer in
Figs.~\ref{fig3}(c)-\ref{fig3}(d) with those in the sandwich
structure in Figs.~\ref{fig3}(e)-\ref{fig3}(f), we see clearly
that borophene receives twice the number of electrons in the
sandwich in comparison to the bilayer structure. Specifically,
$\langle{\Delta}Q\rangle$ in $\epsilon_3$-B almost doubles from
$-0.16$~e/atom in Fig.~\ref{fig3}(c) to $-0.31$~e/atom in
Fig.~\ref{fig3}(e). Similarly, $\langle{\Delta}Q\rangle$ in
$\epsilon_6$-B almost doubles from $-0.21$~e/atom in
Fig.~\ref{fig3}(d) to $-0.41$~e/atom in Fig.~\ref{fig3}(f).

\begin{figure}[h]
\includegraphics[width=1.0\columnwidth]{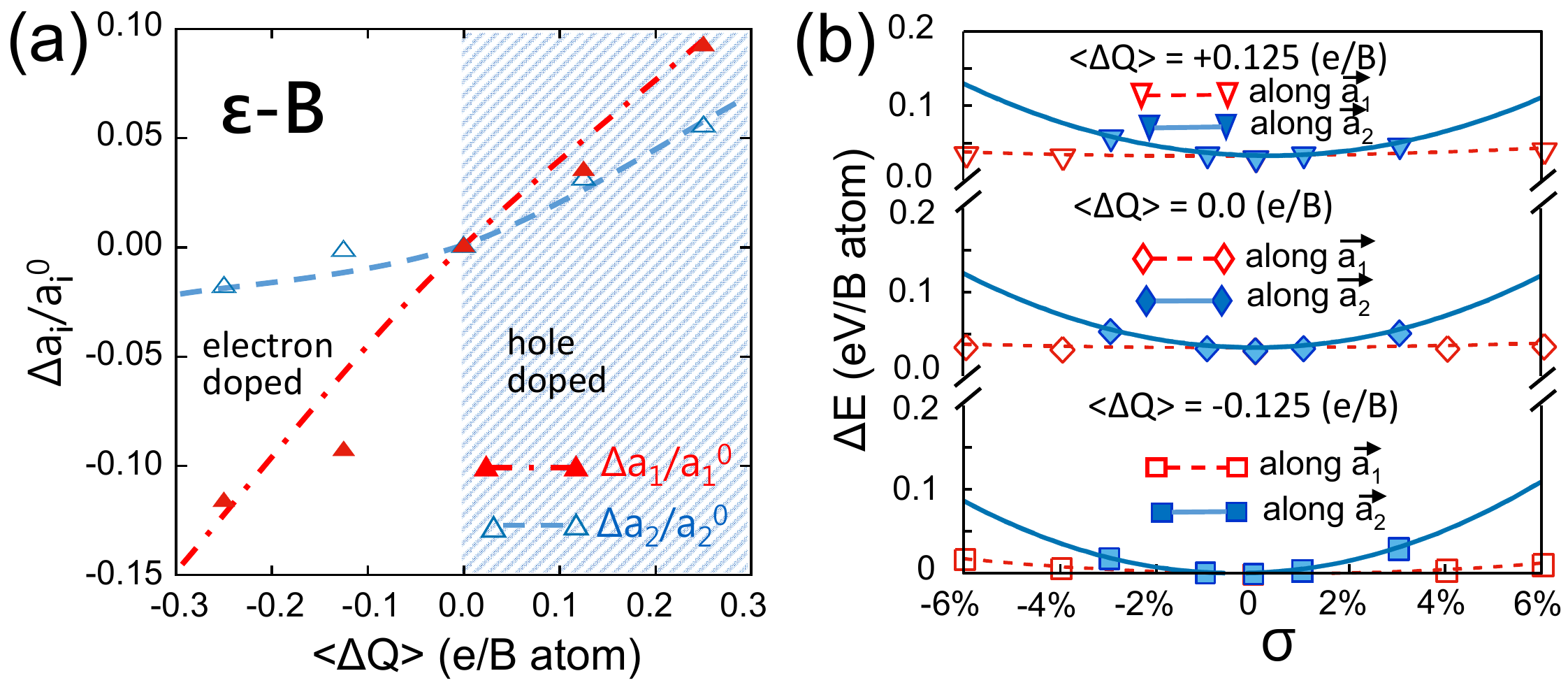}
\caption{Effect of doping level on the equilibrium geometry of
$\epsilon$-B. %
(a) Effect of the net average charge $\langle{\Delta}Q\rangle$ on
the orthogonal lattice constants $a_i$ with $i=1,2$. Plotted are
charge-induced relative changes ${\Delta}a_i/a_i^0$, where $a_i^0$
are the lattice constants in the neutral $\epsilon$-B allotrope. %
(b) Strain energy ${\Delta}E$ as a function of in-layer strain
$\sigma$ at different doping levels $\langle{\Delta}Q\rangle$. %
\label{fig5}}
\end{figure}

\begin{figure*}[t]
\includegraphics[width=1.6\columnwidth]{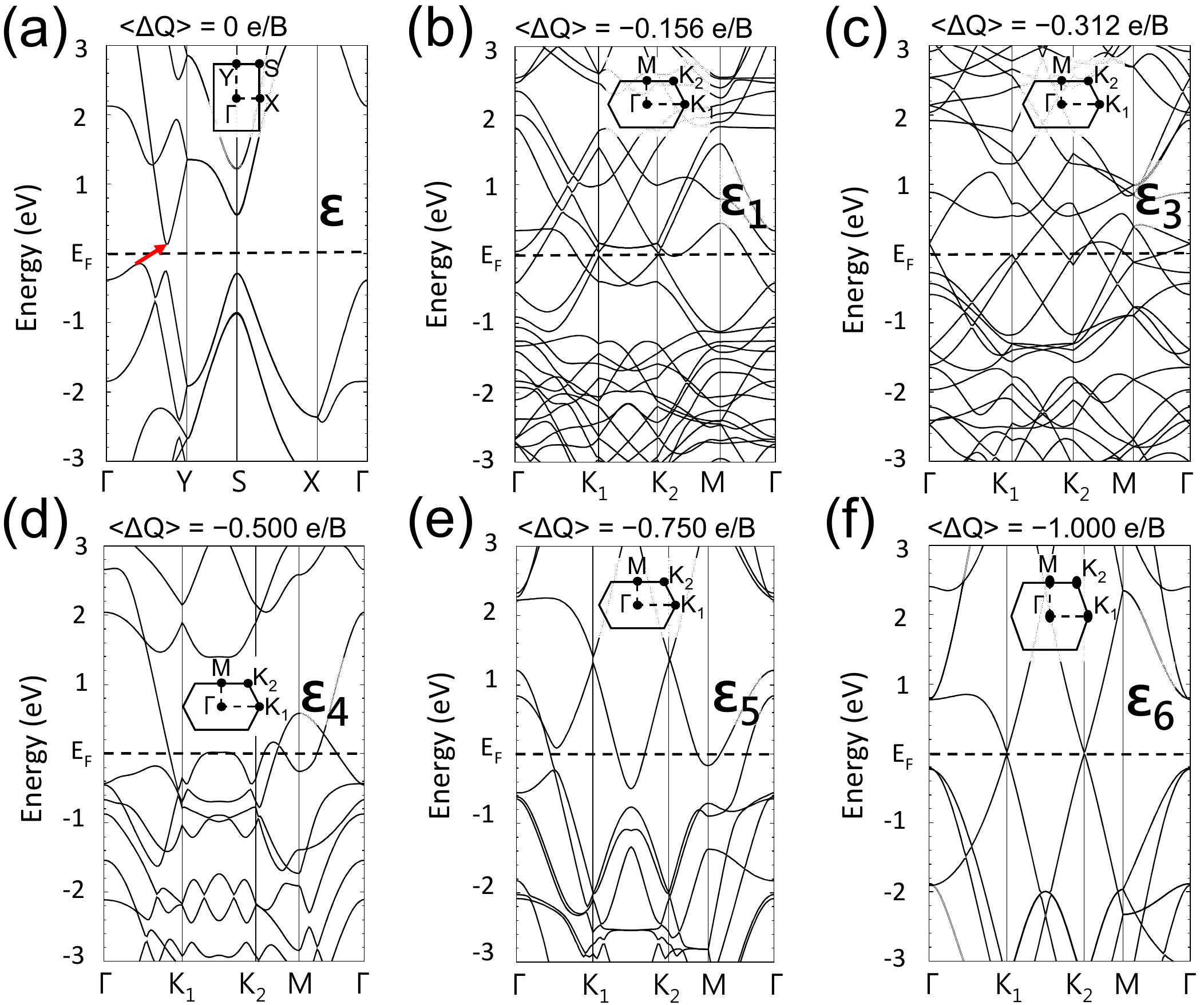}
\caption{Electronic band structure of %
(a) neutral $\epsilon$-B, %
(b) $\epsilon_1$-B with $\langle{\Delta}Q\rangle=-0.156$~e/B, %
(c) $\epsilon_3$-B with $\langle{\Delta}Q\rangle=-0.406$~e/B, %
(d) $\epsilon_4$-B with $\langle{\Delta}Q\rangle=-0.500$~e/B, %
(e) $\epsilon_5$-B with $\langle{\Delta}Q\rangle=-0.750$~e/B, and %
(f) $\epsilon_6$-B with $\langle{\Delta}Q\rangle=-1.000$~e/B, %
calculated using the DFT-PBE functional. The indirect
fundamental band gap is indicated by the red arrow in (a). %
\label{fig6}}
\end{figure*}

We have evaluated the dependence of the total energy on the
interlayer distance $d$ in the Ca$_2$N/B/Ca$_2$N sandwich
structure for the $\epsilon_3$-B and $\epsilon_6$-B allotropes and
present our results in Figs.~\ref{fig4}(a) and \ref{fig4}(b). In
both structures, the optimum interlayer distance
$d{\approx}4.0$~{\AA}.

There is a significant electron accumulation with a maximum at
$d{\approx}3$~{\AA} outside a free-standing Ca$_2$N electride
layer~\cite{Scott16}, which is accommodated by an adjacent
borophene layer. In this case, the net charge on the borophene
layer may be changed by changing the interlayer distance $d$. Our
results for $\langle{\Delta}Q\rangle$ as a function of $d$ are
presented in Fig.~\ref{fig4}(c) for $\epsilon_3$-B and in
Fig.~\ref{fig4}(d) for $\epsilon_6$-B sandwiched in-between two
Ca$_2$N layers. We find it interesting that
$\langle{\Delta}Q\rangle$ decreases almost linearly with
increasing interlayer distance.

At the equilibrium interlayer distance, we find
$\langle{\Delta}Q\rangle{\approx}-0.31$~e/B-atom in the system
with $\epsilon_3$-B and
$\langle{\Delta}Q\rangle{\approx}-0.41$~e/B-atom in the system
with $\epsilon_6$-B. The higher value of $\langle{\Delta}Q\rangle$
in $\epsilon_6$-B is associated with the better ability of this
structure to accept electrons.

It is well known that in GICs, the net charge transferred from
dopant atoms to the graphene layers changes the lattice constant.
This change is relatively small, amounting to~\cite{Nixon69}
${\Delta}a/a{\approx}0.8$\% in KC$_8$ with
$\langle{\Delta}Q\rangle{\approx}-0.125$~e/C-atom. Our
corresponding results for the effect of doping on the lattice
constants in the stable $\epsilon$-B allotrope are shown in
Fig.~\ref{fig5}(a). The lattice changes are anisotropic and larger
than found in graphene. We find that electron doping expands the
lattice more along the softer $\vec{a}_1$ direction than along the
harder $\vec{a}_2$ direction. At the electron doping levels
$\langle{\Delta}Q\rangle{\approx}-0.3$ to $-0.5$~e/B-atom
discussed above for the Ca$_2$N/B/Ca$_2$N heterostructures, the
lattice expansion exceeds 10\%.

The in-layer stiffness of neutral, electron- and hole-doped
$\epsilon$-B is addressed in Fig.~\ref{fig5}(b). We used the
shorthand notation $0$ for the neutral system, %
$-$ for $\langle{\Delta}Q{\rangle}=-0.125$~e/B and %
$+$ for $\langle{\Delta}Q{\rangle}=+0.125$~e/B doping.
Irrespective of doping, the energy change due to in-layer strain
$\sigma$ is much larger along the harder $\vec{a}_2$ direction
than along the softer $\vec{a}_1$ direction. In terms of the 2D
elastic constants~\cite{DT255}, we find for the softer $\vec{a}_1$
direction
$c_{11}(-)=67.99$~N/m, %
$c_{11}(0)=18.70$~N/m, and %
$c_{11}(+)=29.98$~N/m. %
Along the harder $\vec{a}_2$ direction we find
$c_{22}(-)=446.79$~N/m, %
$c_{22}(0)=360.52$~N/m, and %
$c_{22}(+)=338.58$~N/m. %


\subsection*{Electronic structure of borophene allotropes }

The electronic band structure of selected boron 2D allotropes
discussed in this study is shown in Fig.~\ref{fig6}. We should
note that DFT calculations used in this study do not represent the
true quasi-particle band structure and typically underestimate
band gaps. With this fact in mind, we find that, according to the
PBE results in Fig.~\ref{fig6}(a), the neutral $\epsilon$-B
allotrope of Fig.~\ref{fig1}(a) is a semiconductor with a small
indirect band gap of $E_g=0.2$~eV. Whereas stretching along the
soft $\vec{a}_1$ direction by 4{\%} turns $\epsilon$-B into a
direct-gap semiconductor, compressing by 4{\%} along $\vec{a}_1$
causes gap closure. Stretching $\epsilon$-B by 1{\%} along the
hard $\vec{a}_2$ direction turns this allotrope metallic, whereas
compression by 1{\%} changes the its indirect gap to a direct gap.
More details about the electronic structure of strained
$\epsilon$-B are provided in the SI.

We find all doped borophene allotropes to be metallic or
semi-metallic, as shown in Figs.~\ref{fig6}(b)-\ref{fig6}(f).
Inspection of the band structure reveals the formation of a Dirac
cone at K$_1$ and $K_2$ in the honeycomb structures in
$\epsilon_5$-B in Fig.~\ref{fig6}(e) and in $\epsilon_6$-B in
Fig.~\ref{fig6}(f). The Dirac cone appears ${\approx}1.4$~eV above
$E_F$ in $\epsilon_5$-B at the doping level
$\langle{\Delta}Q\rangle=-0.750$~e/B and at $E_F$ in
$\epsilon_6$-B at $\langle{\Delta}Q\rangle=-1.000$~e/B, which
mimics the structure and valence charge of graphitic carbon.

\section*{Discussion}

The majority of the reported stable structures of neutral 2D
borophene were triangular lattices containing arrays of monatomic
vacancies, reflecting the frustrated bonding character of
electron-deficient boron. When considering the effect of excess
charge on the bonding geometry, we observed a transition driven by
increasing net negative charge from structures containing
triangles and higher polygons, depicted in Fig.~\ref{fig1}(a) and
Figs.~\ref{fig2}(a)-\ref{fig2}(c), for
$|\langle{\Delta}Q\rangle|<0.5$~e/B to all-hexagon structures,
depicted in Figs.~\ref{fig2}(c)-\ref{fig2}(f), for
$|\langle{\Delta}Q\rangle|>0.5$~e/B. The highest coordinations
number of six that may be achieved in a network of triangles
reflects to some degree the vain attempt of neutral boron atoms to
satisfy the octet rule. Excess negative charge, with the maximum
value $|\langle{\Delta}Q\rangle|=1.0$~e/B considered here, offers
the ability to satisfy this rule in the honeycomb network of
$\epsilon_6$-B in Fig.~\ref{fig2}(f).

To understand the chemical origin of these structural changes, we
need to inspect the charge redistribution caused by additional
doping, corresponding to the Fukui function for a crystal.
Corresponding results for ${\Delta}\rho$ in selected electron
doped borophene structures are presented in the SI. Complementing
our results presented in Fig.~\ref{fig3} for borophene interacting
with Ca$_2$N, these results suggest that additional electrons are
first accommodated in $p_z$ (or $\pi$) states normal to the
borophene layer, which cause buckling, and then in $\sigma$
states, which reduce the amount of buckling. Starting with the
lightly electron doped $\epsilon_1$-B containing triangles, we
observe an increasing degree of buckling with increasing electron
doping up to $|\langle{\Delta}Q\rangle|{\lesssim}0.5$~e/B. At that
point, the structure changes to the honeycomb structure mimicking
hole-doped graphene. For $|\langle{\Delta}Q\rangle|>0.5$~e/B,
additional excess charge gets increasingly accommodated in
$\sigma$ states, which are much closer to $E_F$ in borophene than
in graphene, causing a reduction of buckling down to zero for
$\langle{\Delta}Q\rangle=-1.0$~e/B.

Assuming that the Ca$_{2}$N electride can transfer up to one
electron per formula unit to a borophene layer, we can expect the
maximum average charge $\langle{\Delta}Q\rangle$ per boron atom in
B/Ca$_{2}$N bilayers to range from
$-0.22$~e in $\epsilon_3$-B/Ca$_2$N to $-0.28$~e/atom in
$\epsilon_6$-B/Ca$_2$N. The transferred charge may be up to twice
as large in Ca$_{2}$N/B/Ca$_{2}$N sandwich structures. These
values are slightly larger, but close to those found in the actual
heterostructures, reported in Fig.~\ref{fig3}. We compared the
charge transfer between monolayers and multilayers of the Ca$_2$N
electride in contact with borophene and found essentially no
difference, as seen in the SI. Thus, the number of Ca$_2$N layers
does not affect the maximum value of $\langle{\Delta}Q\rangle$.

It is not easy to achieve the doping level
$\langle{\Delta}Q\rangle=-1.0$~e/B by contacting an
electronegative material. According to our results for the
Ca$_{2}$N/B/Ca$_{2}$N sandwich structure in Fig.~\ref{fig4}(d),
the Ca$_{2}$N electride can provide only up to ${\approx}0.4$
electrons per boron atom, much less than the desired doping level
of $1$~e/B. An alternative to Ca$_{2}$N is Y$_{2}$C, which can
supply twice as many electrons as Ca$_{2}$N, but is hard to
exfoliate. Assuming maximum charge transfer from Y$_{2}$C to
borophene, B could receive up to $0.8$ electrons in the
Y$_{2}$C/B/Y$_{2}$C heterostructure. Such a large electron
transfer should further augment the Coulomb attraction between
borophene and Y$_{2}$C, thus further deducing the interlayer
distance, as seen in Figs.~\ref{fig3}(a) and \ref{fig3}(b), and
increase the amount of electron transfer, possibly up to
$1.0$~e/B.

We have mentioned electronic structure parallels between the
honeycomb structure of $\epsilon_6$-B carrying one extra electron
per atom and graphene. Even though the system of $\pi$ electrons
near $E_F$ and the Dirac cone in the corner of the Brillouin zone
occur in both systems, there are notable differences between the
systems. In graphene, the top of the $\sigma$-band lies more than
$3$~eV below $E_F$, whereas this energy difference is only
$0.2$~eV in $\epsilon_6$-B. Apparently, the lower core charge of
elemental boron is the main reason, why the $\sigma$ and $\pi$
bands are energetically closer than in graphene. As mentioned
before, this results in an increased role of $\sigma$ states in
electron doped borophene structures.

According to our results in Fig.~\ref{fig5}(a), changing the
doping level $\langle{\Delta}Q\rangle$ from $-0.25$~e to $+0.25$~e
results in a $23$\% increase of the borophene lattice constant
$a_1$. Conversely, we may speculate that changing the lattice
constant should modulate the electron transfer from the electride
to the borophene layer, thus changing the dipole moment normal to
the interface. In that case, an in-plane vibration of the
heterostructure will cause this dipole to oscillate and to emit an
electromagnetic signal of the same frequency.

Clearly, the most drastic effect of electron doping is seen in the
hexagonal borophene structure stabilized in MgB$_2$, which is the
cause of superconductivity in this system. Electron doping may
also be used as a way to change structures in a predictable way.
As an illustration, we find the $\chi_3$-B
structure~\cite{Kehui16}, which has been synthesized at 680~K on
Ag(111), to be very similar to the electron-doped, buckled
$\epsilon_3$-B structure. In that case, a simpler way to fabricate
$\chi_3$-B may consist of bringing borophene in contact with an
electride, allowing it to relax to $\epsilon_3$-B while electron
doped. The final flat structure of $\chi_3$-B may evolve after the
electride has been removed. These two examples illustrate new
possibilities of using doping to change the structure of boron.
Even though doping may have a lesser effect on the equilibrium
structure of other systems, our findings about the interplay
between structure and excess charge have a general validity.

\section*{Summary and Conclusions}


In summary, we have studied the effect of electron doping on the
bonding character and stability of two-dimensional (2D) structures
of elemental boron, called borophene, which is known to form many
stable allotropes. Our {\em ab initio} calculations for the
neutral system have revealed previously unknown stable 2D
$\epsilon$-B and $\omega$-B structures in addition to previously
reported triangular lattices with monatomic vacancies. We found
that the chemical bonding characteristic in this and other boron
structures is strongly affected by extra charge, which is first
accommodated in the $\pi$ and subsequently in the $\sigma$ network
of electrons. Beyond a critical degree of electron doping, the
most stable allotrope was found to change from $\epsilon$-B
containing triangles and higher polygons to a buckled honeycomb
structure. Additional electron doping, mimicking a transformation
of boron to carbon, causes a gradual decrease in the degree of
buckling of the honeycomb lattice that can be interpreted as
piezoelectric response. We also found that net electron doping can
be achieved by placing borophene in direct contact with layered
electrides such as Ca$_{2}$N. We found that electron doping can be
doubled to ${\approx}0.4$~e/B atom by changing from the
B/Ca$_{2}$N bilayer to the Ca$_{2}$N/B/Ca$_{2}$N sandwich
geometry.

\section*{Computational Techniques}

Our calculations of the stability, equilibrium structure and
electronic structure have been performed using density functional
theory (DFT) as implemented in the {\textsc{SIESTA}}~\cite{SIESTA}
code. Periodic boundary conditions have been used throughout the
study, with monolayers represented by a periodic array of slabs
separated by a 30~{\AA} thick vacuum region. We used the
Perdew-Burke-Ernzerhof (PBE)~\cite{PBE} exchange-correlation
functional. The {\textsc{SIESTA}} calculations used
norm-conserving Troullier-Martins
pseudopotentials~\cite{Troullier91}, a double-$\zeta$ basis
including polarization orbitals, and a mesh cutoff energy of
$250$~Ry to determine the self-consistent charge density, which
provided us with a precision in total energy of
${\lesssim}2$~meV/atom. The reciprocal space has been sampled by a
fine grid~\cite{Monkhorst-Pack76} of $4{\times}4$~$k$-points in
the 2D Brillouin zones (BZ) of the primitive unit cells of neutral
and doped borophene containing $32$ atoms, $3{\times}3$~$k$-points
in the BZ of supercell of heterostructure of $Ca_{2}N$ and
$\epsilon_6$, and $3{\times}1$~$k$-points in the BZ of supercell
of heterostructure of $Ca_{2}N$ and $\epsilon_3$. Geometries have
been optimized using the conjugate gradient (CG)
method~\cite{CGmethod}, until none of the residual
Hellmann-Feynman forces exceeded $10^{-2}$~eV/{\AA}.

\begin{suppinfo}
Additional results are presented for the characterization of
borophene layers under conditions not covered in the main
manuscript. These include optimum geometries and charge
distribution in free-standing electron doped borophene as well as
in vertical heterostructures of borophene and Ca$_2$N bilayers.
Also presented are electronic band structures of systems described
in the main manuscript, results characterizing the effect of
strain on the electronic band structure, and charge density
differences corresponding to the Fukui function for selected doped
borophene allotropes.
\end{suppinfo}
\quad\par

{\noindent\bf Author Information}\\

{\noindent\bf Corresponding Author}\\
$^*$E-mail: {\tt tomanek@pa.msu.edu}

{\noindent\bf Notes}\\
The authors declare no competing financial interest.

\begin{acknowledgement}
The authors acknowledge financial support by the NSF/AFOSR EFRI
2-DARE grant number EFMA-1433459. We thank Xianqing Lin, Andrii
Kyrylchuk and Rick Becker for many enriching discussions.
Computational resources have been provided by the Michigan State
University High Performance Computing Center.
\end{acknowledgement}

%

\providecommand{\latin}[1]{#1} \makeatletter \providecommand{\doi}
  {\begingroup\let\do\@makeother\dospecials
  \catcode`\{=1 \catcode`\}=2 \doi@aux}
\providecommand{\doi@aux}[1]{\endgroup\texttt{#1}} \makeatother
\providecommand*\mcitethebibliography{\thebibliography} \csname
@ifundefined\endcsname{endmcitethebibliography}
  {\let\endmcitethebibliography\endthebibliography}{}

\end{document}